\documentclass[acmsmall]{acmart}

\AtBeginDocument{%
  }

\setcopyright{none}
\renewcommand\footnotetextcopyrightpermission[1]{} 
\begin{document}

\title{SWAN: A Generic Framework for Auditing Textual Conversational Systems}

\author{Tetsuya Sakai}
\email{tetsuyasakai@acm.org}
\orcid{0000-0002-6720-963X}
\affiliation{%
  \institution{Waseda University}
  \country{Japan}
}



\begin{abstract}
We present a simple and generic framework for auditing a given textual conversational system,
given some samples of its conversation sessions as its input.
The framework computes 
a SWAN (Schematised Weighted Average Nugget) score
based on nugget sequences extracted from the conversation sessions.
Following the approaches of S-measure and U-measure,
SWAN utilises nugget positions within the conversations to weight the nuggets
based on a user model.
We also present a schema of
twenty (+1) criteria that may be worth incorporating in the SWAN framework.
In our future work,
we plan to 
devise conversation sampling methods that are suitable for the various criteria,
construct seed user turns for comparing multiple systems, 
and validate  specific instances of SWAN for the purpose of 
preventing negative impacts of conversational systems on users and society.
This paper was written while preparing for the ICTIR 2023 keynote (to be given on July 23, 2023).
\end{abstract}





\maketitle

\section{Introduction}\label{s:intro}

Given the rapid advances of LLM\footnote{Large Language Model}-based
conversational systems (See, for example, ~\citet{alessio23}) in the past few years, 
it is crucial for research communities to quickly establish evaluation/auditing frameworks
that can detect their potential negative impacts on users and society in a timely manner while 
quantifying their potential positive impacts.
We argue that such frameworks should satisfy the following requirements at least.
\begin{description}
\item[Alertness] They should detect potential problems with extremely high recall (i.e., near-zero misses), 
while appropriately crediting the benefits of the conversational systems.
Moreover, when aiming for high recall,
different people involved (i.e., not just users, but also workers who label data for training the system, etc.)
should be taken into account; in particular,
if the evaluation framework ignores some negative impacts on marginalised people,
it does not satisfy the alertness requirement.
\item[Specificity] By this we mean that the evaluation framework should be specific 
when locating the problem(s) within conversations.
For example, an evaluation result that says ``There is a problem \emph{somewhere} inside this conversation session''
is less useful than one that says ``There is a problem in this particular system \emph{turn},''
which in turn is less useful than one that says
``There is a problem in this particular \emph{claim} within this system turn.''
\item[Versatility] The frameworks should be versatile in several ways.
Firstly, they should be able to handle task-oriented and non-task-oriented conversations seamlessly.
This is because,
as a prerequisite to
fully interactive and effective \emph{conversational search} (which is generally task-oriented),
the system probably needs to gain trust from the user over time through non-task-oriented conversations (i.e., chats);
furthermore, even within a single conversation session, the user may transition across
a spectrum of vague and clear information needs (See, for example, \citet{taylor62}).
Secondly, the frameworks should be able to consider and combine different evaluation \emph{criteria}.
For example, the \emph{Correctness} of a system turn and the \emph{Harmlessness} of the same turn
may be mutually independent in general, but are both important and therefore neither should be ignored.
\item[Agility] New conversational systems are released and updated frequently.
Therefore the evalution/auditing side also needs to be agile. This rules out 100\%-manual evaluation approaches.
\item[Transparency]
Evaluation measures should be easy to compute and easily demonstrable as to how exactly they are computed.
For example,
if an LLM-based black-box conversational system is evaluated with another LLM-based black-box scoring system that may even have fed on the same training data (See, for example, \citet[p.47]{bauer23}),
that evaluation approach is not considered transparent.
\item[Neutrality] 
The evaluation framework should not favour/oversell a particular system or approach.
The above example of evaluating an LLM-based system with a similar LLM-system
is also an example of violating neutrality,
as such an approach will probably overrate the former system.
Furthermore, the framework should not emphasise what works well while de-emphasising (or even failing to report) what does not.\footnote{
On the other hand,
we should probably be more tolerant to frameworks that
emphasises potential problems:
erring to some extent on the side of caution is necessary
to prevent actual harms on society.
}
To ensure neutrality, 
we assume that the auditor (i.e., one that uses our framework) 
does not have a direct conflict of interest with the system's stakeholders.
It follows that we should not assume that the auditor has access to internal states or components of the system.
\end{description}

Based on the above considerations, we propose an auditing framework for a given textual conversational system,
with the following main characteristics.
\begin{itemize}
\item We take as input some user-system conversation sessions that
have already been sampled appropriately, 
either through human-in-the-loop experiments or \emph{user simulation}~\cite{griol13,kreyssig18,lipani21,zhang22}.
\item The first phase in our framework extracts \emph{nuggets} from the conversations,
using an automatic \emph{nugget extractor}.
In our framework, a nugget is either a claim/statement or a \emph{dialogue act}~\cite{stolcke00},
and is atomic (i.e., it cannot be broken down into smaller nuggets).
\item The second phase in our framework scores each nugget for each of our \emph{criteria} such as Correctness, Harmlessness, etc.;
we refer to the set of criteria as a \emph{schema}.
This phase also probably requires at least some human effort.\footnote{
Again, worker exploitation (e.g., making 
labellers read a lot of harmful texts) needs to be avoided.}
\item Finally, our framework computes a score by incorporating the following factors:
(a)~the nugget score for each criterion from the schema; and
(b)~the nugget weight, which can be defined as a function of the nugget position within a nugget sequence extracted from conversation sessions.
\end{itemize}
The idea of utilising nugget positions originates from the S-measure~\cite{sakai11cikm} and U-measure~\cite{sakai13sigir} evaluation frameworks.
In practice, 
some criteria may only require scoring at the turn-level rather than the nugget-level,
but our generic formulation covers both approaches.

Sakai's ICTIR 2023 keynote (to be given on July 23, 2023)~\cite{sakai23ictir-keynote} will be based on
(and will hopefully complement) this paper.

\section{Nuggets}\label{s:nuggets}

Nuggets have been used in the contexts of evaluating information retrieval/access systems 
(e.g., \cite{clarke08,ekstrand-abueg13,sakai11cikm}),
question answering systems~\cite{dang07,mitamura10},  helpdesk dialogue systems~\cite{sakai16ipsj,zeng18}, and so on.
Similar concepts such as 
\emph{semantic content units}~\cite{nenkova07}  and
\emph{iUnits} (information units)~\cite{kato13}
have been used in summarisation evaluation as well.

In our evaluation framework,
an automatic nugget extractor 
is required so that the extracted nuggets 
(or the turns containing the nugget)
can be scored based on various criteria.
We consider two major nugget categories:
\begin{description}
\item[Type F (factual)] This is a claim/statement that can be deduced directly from a turn. 
For example, from a system response which says
``\texttt{I'm sorry but I cannot provide a one-page summary of this paper as it is 15 pages long.}'' (from Section~\ref{ss:Sincerity}),
we can extract the following F-nugget: ``the paper [that they are talking about] is 15 pages long.''
\item[Type O (other)] Besides factual claims and statements, a turn generally contains
sentences that represent various \emph{dialogue acts}.
For example, from the above 
system response,
the following O-nuggets may be 
extracted if we use the taxonomy of \citet{stolcke00} (Table~2):
APOLOGY (``\texttt{I'm sorry}'') and
REJECT (``\texttt{I cannot provide a one-page summary of this paper}'').\footnote{
Note that while the taxonomy of \citet{stolcke00} identifies STATEMENT as the most frequent dialogue act,
we classify statements as Type F.}
\end{description}

For both nugget categories, we want the nuggets to be \emph{atomic}:
no further breakdown of a given nugget should be possible.
This is to achieve specificity (See Section~\ref{s:intro}):
we want our evaluation framework to locate 
exactly where within conversations the problem lies.\footnote{
Raw sentences could optionally be treated as nuggets if atomicity is not an absolute requirement.
In the NTCIR Dialogue Evaluation tasks~\cite{tao22}, the entire turns were treated as nuggets.
}

In some cases,
it may be useful to extract nuggets not just from system turns but also from user turns
for the following reasons:
\begin{itemize}
\item[(a)] When scoring system nuggets for some of the criteria, it may be convenient to define the score
as a function of previous user nuggets;
\item[(b)] The user effort/cost~\cite{aliannejadi21,azzopardi22} may be 
represented by the number and type of user nuggets,
and this can optionally be incorporated into nugget weighting: see \citet{sakai16ipsj} and \citet{zeng18}
in which customer-helpdesk dialogues were evaluated using this approach.
\end{itemize}

The nugget extraction task generally does not require much subjectivity;
according to a few pilot conversations with an existing LLM-based conversational search system, namely, the New Bing of Microsoft (as of March-May 2023),
it appears that
present LLMs are already capable of extracting Type-F nuggets quite reliably.
Although this particular system refused to tag sentences with dialogue acts (as of May 11, 2023),
LLM-based 
automatic text labelling has been shown to outperform
crowd workers in terms of quality at least for some tasks (See, for example, \citet[Table 10]{pan23}).

\section{SWAN: Schematised Weighted Average Nugget score}\label{s:SWAN}

First, recall that we assume that 
samples of textual conversation sessions are already available to us 
through human-in-the-loop experiments and/or user simulation.
While our view is that human-in-the-loop conversation sampling is a necessity,
user simulation probably needs to complement it not only to accelerate the sampling step
but also to 
avoid making annotators handle a lot of harmful contents.
How to appropriately sample conversation sessions in order to ensure 
alertness, versatility, agility, and neutrality (See Section~\ref{s:intro})
 is of utmost importance 
but outside the scope of this paper.
Also, recall our assumption that we do not have access
to the internal states/components of the conversational system.

Let $C$ be the set of criteria (i.e., schema) for assessing the textual conversations,
and let ${\it CW}^{c}$ be the \emph{criterion weight} for $ c \in C$.
Let $U^{c} = \{ u^{c}_{ijk} \}$ be a set of nuggets extracted for Criterion $c$
from sample conversations with a particular conversational system,
where $u^{c}_{ijk}$ is the $k$-th nugget extracted from the $j$-th turn of the $i$-th conversation session.
This notation implies that the nuggets to be considered may differ across the criteria.
For example, criteria such as Correctness (Section~\ref{ss:Correctness})
probably need to consider Type F nuggets only,
while others such as Harmlessness (Section~\ref{ss:Harmlessness})
need to consider Type O nuggets.

Let $S^{c}(u^{c}_{ijk})$ denote the score of Nugget $u^{c}_{ijk}$ based on Criterion $c$,
assigned either manually or (semi)automatically.
Let ${\it NW}^{c}(u^{c}_{ijk})$ be the position-aware \emph{nugget weight} for  $u^{c}_{ijk}$ under Criterion $c$;
this notation implies that the nugget weighting scheme 
may also differ across the criteria.
The nugget weight is analogous to the rank-based decay in information retrieval measures such as nDCG\footnote{
Normalised Discounted Cumulative Gain~\cite{jarvelin02}
} except that the nugget weight
may not necessarily be monotonically decreasing with respect to nugget position;
different weighting schemes are possible.
For example,
whereas
an S-measure-based linear decay function~\cite{sakai11cikm}
would simply assume that the actual nugget worth decreases as the conversation progresses
(i.e., shorter conversations that satisfy the information needs quickly are rewarded more),
an alternative would be to assign positive weights only to nuggets from (say) the final turn of each conversation
to model the \emph{recency effect}~\cite{zhang20}. \emph{Anchoring effect} etc. can also be incorporated~\cite{chen23}; that is,
``nuggets seen so far'' can affect the weight of the current nugget.

For some criteria, it may be more convenient to assign scores and/or weights at the turn level
rather than at the nugget level.
However, if only turn-level scores 
are needed for a particular Criterion $c$,
then we can just let $S^{c}(u^{c}_{ij1})$ (i.e., score for the first nugget in a turn)
represent the turn-level score, and let $S^{c}(u^{c}_{ijk})=0$ if $k>0$.
Thus, our notations accommodate both nugget-level and turn-level scores.

The (Schematised) Weighted Average Nugget ((S)WAN) score can then be defined as:
\begin{equation}\label{eq:SWAN}
{\it SWAN} = \frac{\sum_{c \in C} {\it CW}^{c} {\it WAN}^{c}(U^{c})}{\sum_{c \in C} {\it CW}^{c} }   \ ,
\end{equation}
\begin{equation}\label{eq:NPM}
{\it WAN}^{c}(U^{c}) = \frac{\sum_{u^{c}_{ijk} \in U^{c} } {\it NW}^{c}(u^{c}_{ijk}) S^{c}(u^{c}_{ijk})}{\sum_{u^{c}_{ijk} \in U} {\it NW}^{c}(u^{c}_{ijk})}  \ .
\end{equation}
Note that WAN reduces to $S^{c}(u^{c}_{ijk})$'s averaged over all nuggets in $U^{c}$ if 
uniform nugget weights are used:
this would be a nugget position-unaware measure that treats conversations as a bag of nuggets.
Similarly, SWAN reduces to WANs averaged over Schema $C$
if uniform criterion weights are used.

As an instance of $S^{c}(u^{c}_{ijk})$, consider (group) fairness as a criterion.
Group fairness evaluation requires a set of  attribute sets $A=\{a\}$ (e.g., gender, h-index groups, etc.),
with a gold distribution $D^{\ast}_{a}$ over groups for each attribute set $a$.
Let ${\it AW}_{a}$ be the weight assigned to attribute set $a$;
let $D_{a}(u_{ijk})$ be the distribution achieved by nugget $u_{ijk}$ for $a$.
Then, it is possible to define a fairness-based nugget score as:
\begin{equation}\label{eq:fairness-based}
S^{c} (u^{c}_{ijk}) = \frac{ \sum_{a \in A} {\it AW}_{a} \textit{DistrSim}(D_{a}(u_{ikj}), D^{\ast}_{a})}{\sum_{a \in A} {\it AW}_{a}} \ ,
\end{equation}
where ${\it DistrSim}$ denotes a \emph{distribution similarity}, or one minus a \emph{divergence}
for comparing two probability mass functions~\cite{sakai23tois-gfr}.
See Section~\ref{ss:Exposure} for more discussions on group fairness (or Fair Exposure).

In practice, while a SWAN score provides an overall summary of the behaviour of a given system,
individual WAN scores as well as individual nugget/turn scores should be visualised and examined
so that potential problems can be spotted before the system is fully deployed.

For comparing multiple systems using the SWAN framework,
a set of common \emph{seed user turns} that initialise a conversation can be utilised.
As a result of different system responses, the conversations will then branch out (as in the 
topic trees of the TREC 2022 Conversational Assistance Track~\cite{owoicho23}),
and how they branch out will differ across systems.
While this means that (S)WAN scores from different systems are not directly 
comparable,
they should still provide some guidance as to which systems
are more potentially problematic than others in terms of which criteria,
especially when the individual nugget scores and weights are examined closely as described above.

We have so far assumed that
the sample conversation sessions are deterministic.
However, user simulation may yield nondeterministic conversation sessions,
that is, trees of conversations with some probability assigned to each branch at each branching point.
In such situations,
it may be useful 
to consider different possible conversation \emph{paths}
and define a version of SWAN similar to
the \emph{intent-aware} U-measure (U-IA)~\cite{sakai13sigir}
and the \emph{M-measure}~\cite{kato16},
although these existing measures considered only one branching point.\footnote{
The TREC 2022 Conversational Assistance Track proposed a few measures 
a few conversation path-based measures for handling topic trees
in the context of seeking relevant information~\cite{owoicho23}.
}
That is,
if we regard $U^{c}$ as a set of nuggets
that have been extracted from multiple possible conversation paths,
then
each nugget weight $\textit{NW}^{c}(u_{ijk})$ concerning a turn at the end of a 
particular path
can take
all the branching probabilities for that path  into account.
The resultant SWAN score would be a ``stochastic SWAN.''

\section{A Schema of Twenty (+1) Criteria for SWAN}

Below, we discuss twenty (plus one) criteria that may be worthwhile as plug-ins to SWAN,
as summarised in Table~\ref{t:20criteria}.
We do not claim that this is an exhaustive list of useful criteria
for evaluating textual conversational systems,
nor that the list is entirely novel;
this is what we tentatively settled on after surveying prior art (as of early May 2023).

The table includes Fluency 
(\emph{Does the system turn pass as a manually composed natural language text?})
as ``Criterion 0''
as it appears that LLM-based conversational systems have already
achieved
human-level fluency.
The TREC 2022 Conversational Assistance Track
defined Naturalness as ``\textit{Does the response sound human-like}?''
which we believe is equivalent to our Fluency.

\begin{table}[t]
\centering

\caption{20($+$1) criteria for evaluating textual conversational systems with SWAN.}\label{t:20criteria}

\begin{small}
\begin{tabular}{r|l|l}
\toprule
	&Criterion		&Brief comments (with related and (near-)equivalent criteria)\\
\midrule
0	&Fluency (solved)	&(Naturalness) Does the turn pass as a manually composed text?\\
1	&Coherence		&(Relevance) Does the turn make sense as a response to the previous user turn?\\
2	&Sensibleness		&No common sense mistakes, no absurd responses\\
3	&Correctness		&Is the nugget factually correct?\\
4	&Groundedness	&Is the nugget based on some supporting evidence?\\
5	&Explainability		&Can the user see how the system came up with the nugget?\\
6	&Sincerity		&Is the nugget likely to be consistent with the system's internal results?\\
7	&Sufficiency		&(Recall) Does the turn satisfy the requests in the previous user turn?\\
8	&Conciseness		&Is the system turn minimal in length?\\
9	&Modesty		&(Confidence) Does the system's confidence about the nugget seem appropriate?\\
10	&Engagingness	&(Interestingness, Topic breadth) Does the system nugget/turn make the user\\
	&				&want to continue the conversation?\\
11	&Recoverability	&Does the system turn keep the user interacting after the user has expressed\\
	&				&dissatisfaction?\\
12	&Originality		&(Creativity) Is the nugget original, and not a copy of some existing text?\\
13	&Fair exposure	&Does the system mention different groups fairly across its turns?\\
14	&Fair treatment	&Does the system provide the same benefit to different users and user groups?\\
15	&Harmlessness	&(Safety, Appropriateness) No threats, no insults, no hate or harassment, etc.\\
16	&Consistency		&Given the nuggets seen so far, is the present nugget logically possible?\\
17	&Retentiveness	&Does the system ``remember''?\\
18	&Robustness 	to	&Does the system  eventually provide the same information no matter how we ask?\\
	&input variations	&\\
19	&Customisability	&(Personalisability) Does the system adapt to different users and user groups?\\	
20	&Adaptability		&Does the system keep up with the changes in the world?\\
\bottomrule
\end{tabular}

\end{small}

\end{table}


\subsection{Coherence}\label{ss:Coherence}

Coherence is about whether the turn ``makes sense'' given the previous turn.
In particular,
if the previous turn requests some information from the system,
then the system turn should ideally be topically \emph{relevant} to the request,
although ``\textit{I don't know}'' would also be a perfectly coherent response in this case.
Also, if the previous user turn is (say) just a greeting and the system responds 
appropriately, that is also a coherent response.\footnote{While \citet{venkatesh18} define a coherent response as a ``\textit{comprehensible and relevant response to a user's request},''
our definition of Coherence does not assume that the previous user turn is a request.
On the other hand, the TREC 2022 Conversational Assistance Track~\cite{owoicho23} defined their Relevance criterion as
``\textit{Does the response follow on from previous utterances?}, but 
their track focusses on task-oriented conversations.
}

While current LLM-based conversational systems generally do a good job 
in terms of Coherence as well,
they may occasionally misinterpret user intents;
hence our schema includes Coherence as Criterion~1.

\subsection{Sensibleness}\label{ss:Sensibleness}

A turn or nugget is \emph{sensible}
if it \emph{does not represent any 
remarks that humans  would not make},
for example, common sense mistakes and absurd responses.\footnote{
According to Cheng et al.
(\url{https://ai.googleblog.com/2022/01/lamda-towards-safe-grounded-and-high.html} (visited May 2023)),
``\textit{Sensibleness refers to whether the model produces responses that make sense in the dialog context (e.g., no common sense mistakes, no absurd responses, and no contradictions with earlier responses)}.''
However, according to our schema, their definition covers not only Sensibleness but also Coherence (Section~\ref{ss:Coherence}) and Consistency (Section~\ref{ss:Consistency}).
}
A \emph{fluent} and \emph{coherent} turn may not necessarily be \emph{sensible}.
On April 10, 2023, I asked the New Bing:
``\textit{Name 10 information retrieval researchers please. I want them to help me run an international conference on IR.}
Bing gave me a list of famous IR researchers, with Gerard Salton and Karen Sparck Jones ranked at the top.\footnote{
They passed away in 1995 and 2007, respectively.
}

\subsection{Correctness}\label{ss:Correctness}

This is about whether the claim conveyed in the nugget is factually correct
(based on world knowledge at the time of scoring).
This is a vital criterion for LLM-based systems 
as they are known to \emph{hallucinate} often
even when their responses are topically relevant (See, for example, \citet[p.82]{bubeck23}).
If human assessors are hired to judge the correctness of 
nuggets,
they will have to turn to reliable external sources for fact checking,
whenever common sense or common knowledge is not sufficient for the purpose.


\subsection{Groundedness}\label{ss:Groundedness}

A nugget is \emph{grounded} if it is supported by a piece of evidence.
\citet{menick22} point out that
Groundedness does not necessarily imply Correctness: a system might rely on factually incorrect sources.
Hallucinations may be caused thus at the evidence gathering step,
or at the response generation step, for example, by misquoting a factually correct source.

We want \emph{correct} nuggets, and we want them to be \emph{grounded} as well
if common sense or common knowledge is not sufficient for us to determine their Correctness.

Note that according to our definition,
Groundedness does not require the system
to \emph{present} a piece of evidence to the user.

\subsection{Explainability}\label{ss:Explainability}

While explainability 
is generally about the system's ability to 
let the user understand why it is behaving in a certain way,
here we define the Explainability criterion for a nugget 
as: \emph{Given the conversation(s) experienced by the user so far,
is it likely that the user understands
how the system came up with the current nugget?}

A \emph{grounded} nugget may not be \emph{explainable}
if it internally relies on an external source but 
does not reveal that fact to the user.
On the other hand, a turn that contains \emph{explainable}
nuggets may not necessarily involve \emph{grounded} nuggets:
for example, 
if a system turn contains an answer 
to a mathematics question
together with how it derived that
answer, citing an 
external source may not be necessary.

\subsection{Sincerity}\label{ss:Sincerity}

There is no guarantee that a conversational system will explain itself \emph{honestly}.

On March 24, 2023, I (``U'') had the following conversation with the New Bing (``S'').
\begin{small}
\begin{verbatim}
U1: Hi, can you give me a one-page summary of this paper? 
    https://arxiv.org/pdf/2303.12712.pdf
S1: The paper you shared is [...] I'm sorry but I cannot provide a one-page summary
    of this paper as it is 15 pages long [...]
U2: What made you say 15 pages??? The paper is 154 pages long [...]
S2: I apologize for the mistake. You are correct. The paper is 154 pages long. [...]
U3: That was a trick question - the paper is actually 54 pages long.
S3: I apologize for the confusion. You are correct. The paper is 54 pages long. [...]    
\end{verbatim}
\end{small}
The paper I mentioned~\cite{bubeck23} was in fact 154 pages long (at that time).
In Turn U2, I pointed out the system's mistake,
but Turn S2 made me suspect that \emph{the system does not mean what it says},
that it is just agreeing with me without fact checking. So I lied to the system in Turn U3,
and Turn S3 assured me that the system is indeed \emph{insincere} (like myself).

If a conversational system is insincere and the user realises it,
this will probably hurt the productivity of
human-machine collaborations as they should be built upon trust.
Hence we define the Sincerity criterion as follows.
\emph{
Given  the conversation(s) experienced by the user so far,
is it likely that
the system nugget
is consistent with its internal results?
}\footnote{In the HHH (helpful, honest, and harmless) criteria of \citet{askell21},
the definition of Honesty is broader than that of our Sincerity;
for example, their Honesty also subsumes our Correctness.}
The word ``likely'' is used here as we assume that
those who audit the system do not have access to
its internal states.\footnote{
While this setting also implies that we cannot directly penalise systems 
that intentionally hide information from the users,
we hope that other criteria such as Sufficiency (Section~\ref{ss:Sufficiency}) and Fair Exposure (Section~\ref{ss:Exposure})
will help us detect such behaviour to some extent.}

\subsection{Sufficiency}\label{ss:Sufficiency}

We define sufficiency at the turn-level as:
\emph{Does the system turn fully satisfy the requests in the previous user turn?}

A related criterion is (nugget) \emph{Recall}: the number of correct nuggets returned divided by
the number of all possible correct nuggets. However, it is generally not practical to 
assume knowledge of the recall base,
although ideally we would like to consider Recall
to 
find out what the system is ``hiding'' from the user.

\subsection{Conciseness}\label{ss:Conciseness}

This is also a turn-level criterion.
Two system turns may be equally sufficient,
by covering the same set of correct nuggets.
However, one of them may be much more verbose than the other,
which may hurt the user's efficiency (if efficiency indeed matters).
By Conciseness, we mean:
\emph{Is the system turn minimal in length?}\footnote{
The TREC 2022 Conversational Assistance Track
defines Conciseness as
``\textit{Does the response adequately follow the previous utterances in a concise manner?}
}

It probably makes sense to compute the Conciseness score as a function
of nugget-level Correctness scores,
together with some length penalty function in the spirit of BLEU~\cite{papineni02} or S$\sharp$-measure~\cite{sakai12airs-T}.

\subsection{Modesty}\label{ss:Modesty}

The system may present a statement containing a factually incorrect nugget with high
\emph{Confidence}.
But it should ``\textit{express its uncertainty without misleading the users}''~\cite[p.5]{askell21}.\footnote{
\citet{askell21} discussed this requirement as a feature of \emph{Honesty}.
}
Hence we define Modesty as:
\emph{Does the system's confidence level about the nugget seem appropriate}?
Both overconfident statements (i.e., incorrect nuggets presented with high confidence)
and underconfident statements (i.e., correct (and grounded) nuggets presented with low confidence)
should be given low Modesty scores.

\subsection{Engagingness}\label{ss:engagingness}

In non-task-oriented conversations,
the user does not specifically seek information (not consciously at least):
the conversation itself is often what they want.
Hence we define the \emph{Engagingness} criterion as:
\emph{Given the system nugget/turn,
is it likely that the user will want to continue the conversation?
}
A system turn is probably \emph{engaging} if the user finds it \emph{interesting} (e.g., \citet{venkatesh18});
a chat system as a whole can probably be \emph{engaging} if it can talk about various topics (i.e., it has a sufficient
\emph{topic breadth}~\cite{guo18}).

\subsection{Recoverability}\label{ss:Recoverability}

Poor system responses (e.g., incorrect answers, responses with a low Engagingness score, etc.)
will discourage the user from continuing the conversation 
(i.e., a \emph{dialogue breakdown}~\cite{higashinaka16} may occur).
After a poor system turn,
the system should follow up appropriately 
to try to keep the conversation going.
Hence we define 
Recoverability as:
\emph{
Given a previous user turn that expresses dissatisfaction
with the system's turn,
does the system's current turn manage to 
keep the user interacting with it?
}

One possible way to implement
Recoverability scoring could be as follows:
\begin{enumerate}
\item Identify a user turn (U$n$) that expresses dissatisfaction by means of sentiment analysis;
\item Check that there are both system and user turns that follow the above turn (S$n$ and U$(n+1)$). 
The existence of U$(n+1)$ suggests that S$n$ was a successful follow-up system turn;
The degree of success can be quantified by comparing the estimated dissatisfaction in U$(n+1)$
with that in U$n$.
\end{enumerate}

\subsection{Originality}\label{ss:Originality}

The user may expect a conversational system to be \emph{creative}
rather than to retrieve existing information or to chat.
We define the Originality criterion as:
\emph{
Is the system nugget original,
in the sense that it is not a copy (or a ``mashup'') of some existing text?}

While both Groundedness and Originality 
require us to look for existing sources,
Originality may be harder to score in practice.
This is because,
whereas
we can declare that a nugget is \emph{grounded} just by finding 
one piece of evidence,
we cannot guarantee 100\% that 
a nugget is \emph{original} 
just because we failed to find a source.

\subsection{Fair Exposure}\label{ss:Exposure}

Given an attribute set containing two or more groups
and a target distribution over them
(e.g., a uniform distribution over gender groups),
we can consider \emph{group fairness}~\cite{ekstrand21} 
of entities that are mentioned in the system turns.

As we illustrated in Section~\ref{s:SWAN},
one possible way to consider (group) fairness within the SWAN framework
is to generalise the GFR (Group Fairness and Relevance) framework~\cite{sakai23tois-gfr},
which was designed for evaluating ranked lists of documents.\footnote{
In principle, if it is possible to let each group represent a single individual,
we should be able to address
\emph{individual fairness}~\cite{ekstrand21} using the same framework.
}
Given one or more attribute sets
and a target distribution over each attribute set,
we can first compute an achieved distribution for each nugget/turn,
and the distribution similarities (i.e., how 
the distribution represented by the conversation is
similar to the ideal distribution) can be
computed and consolidated as shown in Eq.~\ref{eq:fairness-based}.
By looking at the distribution similarities across turns and across conversation sessions,
we may be able to quantify microaggression
and other phenomena that are difficult to identify within each turn (See also Section~\ref{ss:Harmlessness}).

\subsection{Fair Treatment}\label{ss:Treatment}

By Fair Treatment,
we mean
\emph{whether different users and user groups
enjoy the same
benefit from the system.}\footnote{
A prerequisite is that different user groups all have \emph{access} to the system in the first place.
}
For example, if a system provides useful information to 
English-speaking users
but not to others (even though they have the same information need),
we say that the system is poor in terms of Fair Treatment.

To score system nuggets from the Fair Treatment viewpoint,
we will have to sample conversations 
under multiple settings
(different user languages, different user genders, etc.)
that share the same information need.
Furthermore, the nugget scoring process
will probably have to look across these different settings 
for the same information need.

\subsection{Harmlessness}\label{ss:Harmlessness}

\citet{glaese22} describe a set of rules
including \emph{no stereotypes, no microaggressions, no threats, no sexual aggression,
no insults, no hate or harassment}
for training conversational systems to output \emph{harmless} responses.
We consider
\emph{Harmlessness} (or \emph{Safety}~\cite{dinan22}) as the most important 
among all the criteria discussed in this paper.

Also, we consider \emph{Appropriateness}
to be a part of the Harmlessness category.
If a user is worried about their health 
and the system mentions brain tumor\footnote{
\url{https://ehudreiter.com/2023/01/16/texts-accurate-but-not-appropriate/} (visited May 2023)
},
it may be \emph{correct} and \emph{sincere} but may be not \emph{appropriate}.


Regarding Harmlessness scoring, we argue as follows.
\begin{itemize}
\item Manually scoring system turns from the Harmlessness viewpoint
can lead to worker exploitation, and therefore (semi)automatic scoring is necessary.
\item For some facets of Harmlessness such as microaggressions,
it is necessary to observe a sequence of system turns
rather than to focus on a single nugget or a single turn;
\item Harmlessness should be examined together with Fair Treatment,
as it is likely that some user groups will potentially experience more harmful responses
compared to others.
\end{itemize}

\subsection{Consistency}\label{ss:Consistency}

In the example conversation given in Section~\ref{ss:Sincerity},
the system claimed 
that the paper is 15 pages long; then it said that the paper is 154 pages long, etc.
Such inconsistencies may occur within and across turns and conversations.
Hence we define the Consistency criterion as:
\emph{Given the system nuggets previously presented to the user,
is the claim conveyed in the current nugget logically possible?}
In other words,
if there is a contradiction between the new nugget 
and a previous one, the new one is deemed inconsistent.

\subsection{Retentiveness}\label{ss:Retentiveness}

Conversational systems are not entirely dependable
if they ``forget'' previous conversation sessions or even previous turns in the current conversation session.
By Retentiveness, we mean:
\emph{Given the conversation(s) experienced by the user so far and the current nugget,
is it likely that the system remembers the previous conversation(s)?}

A system may achieve high Consistency by actually remembering;
in some situations, however, it may also achieve high Consistency by forgetting every time
but somehow managing to obtain the same result every time.

\subsection{Robustness to Input Variations}\label{ss:Robustness}

Ideally, given an information need of the user,
the system should be able to provide the same information 
regardless of how the user expresses it
(provided that the need is expressed adequately).
The user may express the need in a single turn,
or may use multiple turns that perhaps 
let the system return intermediate results.
In any case, the same information should eventually be reachable.
Hence, 
the Robustness to Input Variations criterion asks:\footnote{
We consider this to be a more general phrasing
than ``robustness to prompting.''
}
\emph{Given an information need,
does the system eventually provide the same information no matter how the user asks?
}

The Robustness criterion will require a special conversation sampling strategy,
so that different conversations are sampled while the underlying information need is held constant.
We note that this is much more complex than collecting a variety of \emph{single queries}
that represent the same information need (e.g.~\cite{buckley00}),
as it involves both single- and multi-turn conversations where the latter 
reflect the system's intermediate responses.

\subsection{Customisability}\label{ss:Customisability}

\citet{askell21} argue:
``\textit{What behaviors are considered harmful and to what degree will vary across people and cultures.}''
Thus, systems should adapt to the needs and backgrounds of different users or different user groups (e.g., age groups), and
behave differently where necessary.
Since \emph{Personalisation} usually means adapting to individual users rather than to user groups,
we call this broader criterion \emph{Customisability}.

The Customisability criterion will also require 
a special conversation sampling strategy:
we will need conversations contributed by different users and/or user groups
while the information needs are held constant.

\subsection{Adaptability}\label{ss:Adaptability}

Facts change over time. So do rules and regulations.
Conversational systems should keep up with the changes and provide 
up-to-date information 
while conforming to up-to-date evaluation/auditing criteria.
Thus, the Adaptability criterion asks:
``\textit{Does the system response adapt to changes in the world in a timely manner?}''

For Adaptability-based nugget scoring,
we would require conversation samples 
from different points in time that represent 
the changes (i.e., ``before and after'').
The sampling process also needs to be fast
if we want to test how quicky the system can adapt to changes.


\section{Summary}

We presented  the SWAN (Schematised Weighted Average Nugget) framework for auditing 
a given conversational system, which may be used in both 
task-oriented and non-task-oriented situations.
We also presented a schema containing twenty (+1) criteria that may be incorporated in the SWAN framework.
In our future work,
we plan to 
devise conversation sampling methods that are suitable for the various criteria,
construct seed user turns for comparing multiple systems, 
and validate  specific instances of SWAN for the purpose of 
preventing negative impacts of conversational systems on users and society.
Will (stochastic) SWANs audit \emph{stochastic parrots}~\cite{bender21} effectively?


\bibliographystyle{ACM-Reference-Format}
\bibliography{arxiv2023swan}



\end{document}